\hfuzz 2pt
\vbadness 10000
\font\titlefont=cmbx10 scaled\magstep1

\magnification=\magstep1

\null
\vskip 1.5cm
\centerline{\titlefont ENTANGLING OSCILLATORS}
\medskip
\centerline{\titlefont THROUGH ENVIRONMENT NOISE}
\vskip 2.5cm
\centerline{\bf F. Benatti}
\smallskip
\centerline{Dipartimento di Fisica Teorica, Universit\`a di Trieste}
\centerline{Strada Costiera 11, 34014 Trieste, Italy}
\centerline{and}
\centerline{Istituto Nazionale di Fisica Nucleare, Sezione di 
Trieste}
\vskip 1cm
\centerline{\bf R. Floreanini}
\smallskip
\centerline{Istituto Nazionale di Fisica Nucleare, Sezione di 
Trieste}
\centerline{Dipartimento di Fisica Teorica, Universit\`a di Trieste}
\centerline{Strada Costiera 11, 34014 Trieste, Italy}
\vskip 2cm
\centerline{\bf Abstract}
\smallskip
\midinsert
\narrower\narrower\noindent
We consider two independent bosonic oscillators immersed in a
common bath, evolving in time with a completely positive, markovian,
quasi-free (Gaussian) reduced dynamics. We show that an initially 
separated Gaussian state can become entangled as a result of a
purely noisy mechanism. In certain cases,
the dissipative dynamics allows the persistence of these
bath induced quantum correlations even in the asymptotic
equilibrium state.
\endinsert

\vfill\eject

{\bf 1. INTRODUCTION}
\bigskip

When two non-interacting, independent systems are immersed in a common bath,
decoherence effects are expected to arise counteracting any quantum correlation
initially present among the two subsystems [1-5]. 
In certain circumstances however,
the environment can also enhance entanglement. 
Most commonly, this phenomenon is the result of an hamiltonian coupling 
between the two subsystems [6-10] 
generated by the action of the bath; remarkably, it can also occur
in the markovian regime through a non-hamiltonian, purely noisy mechanism [11-22].

This environment induced entanglement generation has been identified so far in models
involving finite-level systems. In this note we shall instead examine the case
of simple infinite dimensional systems following a purely markovian dynamics. 
More specifically, we shall study the behaviour of two independent
bosonic oscillators evolving in time according to a semigroup of completely
positive maps. We shall limit our considerations to the phenomenologically
relevant set of quasi-free (or Gaussian) states [23-29]
and to dynamics that preserve
this set, the so-called quasi-free semigroups, whose properties are well studied
in the literature [30, 31].

After a brief review of the theory of quasi-free dynamical semigroups in relation 
to the present situation, we shall first discuss the possibility 
of entanglement creation at the beginning
of the evolution, in the vicinity of the initial time $t=\,0$.
Using the partial transposition criterion for two-mode Gaussian states, we shall derive
the condition that assures environment induced entanglement generation
as soon as $t>0$.

Although in general entanglement will keep growing also away from the
neighborhood of $t=\,0$, the decoherent, noisy character of the environment will
eventually set in. Whether or not quantum correlations are actually left in the 
asymptotic long time regime is the result of a delicate balance between
entanglement generation and decoherence.
It is remarkable that for certain environments the long time equilibrium
state could indeed retain a non-vanishing entanglement, and an example
of this phenomenon is explicitly presented at the end.
In view of the direct phenomenological relevance of Gaussian states in quantum
optics, we believe that these results might be of help in the design 
and actual realization of quantum devices and circuits.

\vskip 1cm

{\bf 2. GAUSSIAN STATES}
\bigskip

As explained in the introductory remarks, we shall study the dynamics of 
independent oscillators in weak interaction with a large environment,
so that their reduced time evolution can be well represented by a
markovian, completely positive dynamics. Although our considerations
concerning bath-induced entanglement creation will involve two
such oscillators, for sake of generality in this Section and the next will
let their number $n$ to be arbitrary. 
The states of this system will be represented by a density matrix $\rho$, 
{\it i.e.} by a positive hermitian operator, with unit trace, acting on the
bosonic Hilbert space ${\cal H}$. It can be identified with the tensor product
of $n$ independent Fock spaces; they are generated
from the vacuum state through the action of polynomials in the creation $a_i^\dagger$
and annihilation $a_i$, $i=1, 2,\ldots, n$, operators 
pertaining to the single oscillators. These operators obey the standard
bosonic oscillator algebra: $[a_i,\ a^\dagger_j]=\delta_{ij}$,
$[a_i^\dagger,\ a^\dagger_j]=\,0=[a_i,\ a_j]$.

In dealing with this algebra, it is useful to 
adopt the holomorphic representation \hbox{[32-35]};
it allows working with explicit expressions for both states and operators,
including the density matrix $\rho$.
In this formulation, the elements $|\psi\rangle$ of the bosonic Hilbert
space ${\cal H}$ are represented by holomorphic functions $\psi(\bar z)$
of the set of $n$ complex variables 
$\bar z=(\bar z_1,\bar z_2,\ldots, \bar z_n)$, with inner product:%
\footnote{$^\dagger$}{Here and in the following we use the conventions
of Ref.[32]}
$$
\langle\phi|\psi\rangle=\int \psi^*(z)\, \phi(\bar z)\
e^{-\sum_i\bar z_i z_i}\ \Pi_i\, {\rm d}\bar z_i\, {\rm d}z_i\ ,
\eqno(2.1)
$$
where $*$ signifies complex conjugation.
To every operator $\cal O$ acting on ${\cal H}$ there corresponds a kernel
${\cal O}(\bar z, z)$ of $2n$ independent complex variables 
$\bar z=(\bar z_1,\bar z_2,\ldots, \bar z_n)$ and 
$z=(z_1,z_2,\ldots, z_n)$,
such that for the state $|\phi\rangle={\cal O}\,|\psi\rangle$ one finds
the representation:
$$
\phi(\bar z)=\int {\cal O}(\bar z,  w)\, \psi(\bar w)\
e^{-\sum_i\bar w_i w_i}\ \Pi_i\, {\rm d}\bar w_i\, {\rm d}w_i\ .
\eqno(2.2)
$$
In particular, the creation and annihilation operators, when acting on
a state $|\psi\rangle$, are realized by multiplication and
differentiation by the variables $\bar z_i$:
$$
a_i^\dagger |\psi\rangle \rightarrow \bar z_i\, \psi(\bar z)\ ,\qquad
a_i |\psi\rangle \rightarrow {\partial\over\partial\bar z_i} \psi(\bar z)\ ,
\eqno(2.3)
$$
while the identity operator is represented by $\exp({\sum _i\bar z_i  z_i})$.

Among all density matrices $\rho$ for the $n$ oscillators,
the so-called quasi-free or Gaussian states are of particular interest:
they can be easily produced in experiments
in quantum optics [27-29]. In the holomorphic formulation, they can be defined
as possessing a representation kernel $\rho(\bar z, z)$
of generic Gaussian form:
$$
\rho(\bar z,  z)=\sqrt{\cal N}\
\exp\bigg[ -{1\over2}\, {\rm\bf z}^T\cdot\, {\rm\bf G}^{-1}\cdot\, 
{\rm\bf z} +\sum_i\bar z_i z_i\bigg]\ ,
\eqno(2.4)
$$
where ${\rm\bf z}^T$ is a $2n$-dimensional row vector
with components 
$(z_1,z_2,\ldots,z_n,\bar z_1,\bar z_2,\ldots,\bar z_n)$,
while ${\rm\bf z}$ is its corresponding column vector ($T$ signifies
matrix transposition and $\cdot$ matrix multiplication). 
The covariance ${\rm\bf G}$ is
$2n \times 2n$ matrix; with respect to the natural decomposition
${\rm\bf z}^T\equiv (z\ \bar z)$ into the set of variables
$z_i$ and $\bar z_i$, it can be conveniently parametrized
as
$$
{\rm\bf G}=\left(\matrix{\hat\alpha & \hat\beta\cr
                          \hat\beta^T & \hat\alpha^*}\right)\ ,
\eqno(2.5)
$$
where $\hat\alpha$ and $\hat\beta$ are $n \times n$ matrices,
with $\hat\alpha$ symmetric and $\hat\beta$ hermitian to assure
the hermiticity of $\rho$. They represent the averages
in the given state $\rho$ of quadratic operators in
the creation and annihilation operators:
$$
\eqalign{
&\langle a_i\, a_j\rangle\equiv {\rm Tr}\big[a_i\, a_j\, \rho\big]=
\int {\partial^2\over\partial{\bar z}_i\partial\bar z_j}\big[\rho(\bar z, z)\big]\ 
{\rm e}^{-\sum_k\bar z_k z_k}\ \Pi_k\, {\rm d} \bar z_k\, {\rm d}z_k=\hat\alpha_{ij}\ ,\cr
&\langle a^\dagger_i\, a^\dagger_j\rangle\equiv {\rm Tr}\big[a^\dagger_i\,a^\dagger_j\, 
\rho\big]=
\int {\bar z}_j \bar z_j\, \rho(\bar z, z)\ 
{\rm e}^{-\sum_k\bar z_k z_k}\ \Pi_k\, {\rm d} \bar z_k\, {\rm d}z_k=\hat\alpha^*_{ij}\ ,\cr
&\langle a_i\, a^\dagger_j\rangle
\equiv {\rm Tr}\big[a_i\, a^\dagger_j\,\rho\big]=
\int {\partial\over\partial{\bar z}_i}\big[\bar z_j\, \rho(\bar z, z)\big]\ 
{\rm e}^{-\sum_k\bar z_k z_k}\ \Pi_k\, {\rm d} \bar z_k\, {\rm d}z_k=\hat\beta_{ij}\ .}
\eqno(2.6)
$$
As a result of the last relation, the matrix $\hat\beta$ turns out to be non-negative.
For simplicity, in writing (2.4) we have assumed
$\langle a_i^\dagger\rangle=\langle a_i\rangle=\,0$; this
condition can be easily released starting with a more general Ansatz
for $\rho(\bar z, z)$, containing in the exponential also linear 
terms in $z_i$ and $\bar z_i$; it will not be needed for the considerations
that follow.
The trace condition,
$$
{\rm Tr}[\rho]=\int \rho(\bar z, z)\, {\rm e}^{-\sum_i\bar z_i z_i}\
\Pi_i\, {\rm d} \bar z_i\, {\rm d}z_i=1\ ,
\eqno(2.7) 
$$
further fixes the normalization constant, ${\cal N}={\rm det}({\rm\bf G})$,
provided the previous integral makes sense [32]. Indeed, convergence of (2.7), as well as of
the integrals in (2.6), put further constraints on the entries 
of the covariance matrix ${\rm\bf G}$, or equivalently on those of $\hat\alpha$
and $\hat\beta$.
As similarly done with the vector ${\rm\bf z}$, 
let us collect the $2n$
annihilation and creation operators $a_i$ and $a_i^\dagger$ into the column vector
${\rm\bf a}$ and its hermitan conjugate row vector
${\rm\bf a}^\dagger\equiv (a_i^\dagger\ a_i)
=(a_1^\dagger,a_2^\dagger,\ldots, a_n^\dagger,a_1,a_2,\ldots,a_n)$.
Then, necessarily, the following $2n \times 2n$ matrix of bilinear expectation values
results non negative:
$$
\langle {\rm\bf a}_\mu\ {\rm\bf a}_\nu^\dagger \rangle
\equiv{\rm Tr}\big[{\rm\bf a}_\mu\ {\rm\bf a}_\nu^\dagger\ \rho\big] \geq 0\ ,
\qquad \mu, \nu=1,2,\ldots, 2n\ .
\eqno(2.8)
$$
It turns out that (2.8) is also a sufficient condition for the expression in (2.4)
to represent a physical state [23, 36-38].

It is customary to rewrite this condition in terms of the $2n \times 2n$ matrix 
$\rm\bf V$ of symmetric bilinears:
$$
{\rm\bf V}_{\mu\nu}={1\over 2} \Big\langle \big\{{\rm\bf a}_\mu\, ,
\ {\rm\bf a}_\nu^\dagger \big\}\Big\rangle\equiv
{1\over2}{\rm Tr}\Big[\big({\rm\bf a}_\mu {\rm\bf a}_\nu^\dagger
+{\rm\bf a}_\nu^\dagger {\rm\bf a}_\mu\big)\, \rho\Big]\ ;
\eqno(2.9)
$$
in terms of the natural decomposition of
${\rm\bf a}^\dagger\equiv (a_i^\dagger\ a_i)$
into the set of all creation and annihilation operators, it can be explicitly
written as:
$$
{\rm\bf V}=\left(\matrix{\hat\beta & \hat\alpha\cr
                          \hat\alpha^* & \hat\beta^T}\right) -{{\bf 1}\over 2}\ .
\eqno(2.10)
$$
Introducing also the $2n \times 2n$ matrix of commutators
$$
{\bf \Sigma}_{\mu\nu}\equiv\Big\langle\big[{\rm\bf a}_\mu\, ,
\ {\rm\bf a}_\nu^\dagger \big]\Big\rangle=
\left(\matrix{1 & 0\cr
              0 & -1}\right)\ ,
\eqno(2.11)
$$
one can finally express the positivity condition (2.8) as:
$$
{\rm\bf V} + { {\bf \Sigma}\over 2 }\geq 0\ .
\eqno(2.12)
$$
In the following, we shall limit our considerations to the set
of Gaussian states, {\it i.e.} to the states $\rho$ represented
by kernels of the form (2.4) satisfying the condition (2.8),
or equivalently (2.12).

A particularly important class of Gaussian states are the pure ones.
In the holomorphic representation they are described by 
properly normalized Gaussian
functions of the variables $\bar z_i$:
$$
\Psi_\Omega(\bar z)={\rm det}^{1/4}\big(1-\Omega^*\Omega\big)\ 
\exp\bigg[{-{1\over2}\sum_{i,j=1}^n \bar z_i \Omega_{ij} \bar z_j}\bigg]\ ,
\eqno(2.13)
$$
with $\Omega$ a complex, symmetric matrix, such that $|\Omega|\leq 1$
to guarantee norm convergence. The corresponding kernel
$\rho(\bar z, z)\equiv \Psi_\Omega(\bar z)\,\Psi_\Omega^*(z)$ can be cast
in the form (2.4), with submatrix coefficients $\hat\alpha=-\Omega(1-\Omega^*\Omega)^{-1}$ 
and $\hat\beta=(1-\Omega^*\Omega)^{-1}$, respectively.

\vskip 1cm

{\bf 3. QUASI-FREE QUANTUM DYNAMICAL SEMIGROUPS}
\bigskip

Our analysis is based on the
assumption that the time evolution of the set of independent oscillators
immersed in the common bath be markovian and given by a quantum dynamical semigroup;
this is a completely positive, trace preserving, one parameter
family of linear maps, acting on the set of density matrices $\rho$ representing the
oscillator states.
These maps are generated by equations of the following Kossakowski-Lindblad form
[39-41]:
$$
{\partial\rho(t)\over \partial t}={\cal L}[\rho(t)]\equiv
-i\big[H ,\rho(t)\big] + L[\rho(t)]\ ,
\eqno(3.1)
$$
with $H$ and effective hamiltonian and $L$ a dissipative piece,
that can be abstractly written as
$$
L[\rho]=\sum_k \bigg(L_k\,\rho\, L^\dagger_k
-{1\over2}\Big\{L^\dagger_k L_k,\rho\Big\}\bigg)\ 
\eqno(3.2)
$$
in terms of a collection of suitable, well-behaved operators $L_k$.

Being interested in the set of Gaussian states, we would like to characterize
those quantum dynamical semigroups that preserve that set.
Although already discussed using abstract, mathematically rigorous techniques [30, 31], 
this problem has a simple
and direct solution in the holomorphic representation: the semigroup generated
by (3.1) leaves the form (2.4) for $\rho$ invariant provided
$L_k$ is a linear and $H$ a quadratic combination 
of the creation and annihilation operators.
Indeed, only in this case the r.h.s. of (3.1) results quadratic in
the variables $\bar z_i$ and their derivatives when acting on
the representation kernel $\rho(\bar z, z)$, thus preserving
its generic form (2.4); as a consequence, the evolution equation (3.1)
reduces to a linear, differential equation for the entries $\hat\alpha$ and $\hat\beta$
of the covariance matrix
$\rm\bf G$ (given explicitly in (3.7) below).

In view of this result,
the effective hamiltonian will be taken to have the generic
quadratic form
$$
H={1\over2}\sum_{i,j=1}^n \hat\omega_{ij}\, \big\{ a_i^\dagger,a_j\big\}\ ,
\eqno(3.3)
$$
whith the matrix $\hat\omega$ hermitian and positive. Note that terms
containing $a_i a_j$ and their hermitian conjugate can be eliminated
by a suitable unitary canonical transformation \hbox{[32, 34, 35]}, 
and therefore do not appear
in (3.2). Similarly, the most general allowed dissipative term takes
the following form:
$$
\eqalign{
L[\rho]={1\over2}\sum_{i,j=1}^n & \bigg\{\hat\eta_{ij}\Big(\big[a_j\rho, a_i^\dagger\big]
+\big[a_j,\rho a_i^\dagger\big]\Big)
+\hat\sigma_{ij}\Big(\big[a_j^\dagger\rho, a_i\big]+ \big[a_j^\dagger,\rho a_i\big]\Big)\cr
&+{\hat\lambda}_{ij}\Big(\big[a_j\rho, a_i\big] +\big[a_j,\rho a_i\big]\Big)
+{\hat\lambda}_{ji}^*\Big(\big[a_j^\dagger\rho, a_i^\dagger\big] 
+\big[a_j^\dagger,\rho a_i^\dagger\big]\Big)\bigg\}\ ,
}
\eqno(3.4)
$$
where the $n \times n$ coefficients matrices $\hat\eta$, $\hat\sigma$ and $\hat\lambda$
encode the physical properties of the environment and can be expressed
in terms of the Fourier transform of the correlation functions
in the bath [1-5]. 

These parameters are not completely arbitrary. 
First of all, $\hat\eta$ and $\hat\sigma$ need to be
hermitian in order to comply with the hermiticity preserving requirement 
of the generated semigroup. In addition,
the request of complete positivity gives rise to further constraints. 
Indeed, by using the $2n$-dimensional vectors
${\rm\bf a}$ and ${\rm\bf a}^\dagger$ introduced in the previous section,
the dissipative term in (3.4) can be recast in compact form as:
$$
L[\rho]=\sum_{\mu,\nu=1}^{2n} {\rm\bf C}_{\mu\nu} \bigg(
{\rm\bf a}_\nu\, \rho\, {\rm\bf a}_\mu^\dagger
-{1\over2} \Big\{ {\rm\bf a}_\mu^\dagger\, {\rm\bf a}_\nu\, ,\, \rho\Big\}
\bigg)\ ,
\eqno(3.5)
$$
where the bath coefficients are now embedded in the $2n\times 2n$
Kossakowski matrix $\rm\bf C$. 
It is well known that the requirement of complete positivity of the dynamics 
generated by a dissipative term $L[\rho]$ in the form (3.5)
is equivalent to the positivity of the Kossakowski matrix [39-41];
in the present case, using the block decomposition 
introduced before in writing (2.10), 
this condition explicitly reads:
$$
{\rm\bf C}=\left(\matrix{\hat\eta & \hat\lambda^\dagger\cr
                         \hat\lambda & \hat\sigma}\right)\geq0\ .
\eqno(3.6)
$$

The finite-time evolution maps obtained from the equation (3.1), with hamiltonian
as in (3.3) and dissipative term as in (3.5), with ${\rm\bf C}\geq 0$,
are known in the literature as quasi-free quantum dynamical semigroups [30, 31]:
as already observed, they are characterized by the property of transforming
the set of quasi-free (Gaussian) states into itself.

Inserting the general Ansatz (2.4) for the kernel $\rho(\bar z, z;t)$
in the evolution equation (3.1), with (3.3)
and (3.4), and using the prescriptions (2.3), one easily derives
the equation obeyed by the time-dependent covariance ${\rm\bf G}(t)$,
or equivalently those for the submatrices $\hat\alpha$ and $\hat\beta$.
It can be more conveniently rewritten as an equation for
the symmetric covariance matrix $\rm\bf V$ introduced in (2.9):
$$
\partial_t {\rm\bf V}(t)= {\rm\bf A}^\dagger\cdot {\rm\bf V}(t)
+{\rm\bf V}(t)\cdot {\rm\bf A}
+ {\rm\bf B}\ .
\eqno(3.7)
$$
The $2n \times 2n$ matrices ${\rm\bf A}$ and ${\rm\bf B}$ contain the
dependence on the hamiltonian $\hat\omega_{ij}$ and dissipative coefficients
$\hat\eta_{ij}$, $\hat\sigma_{ij}$, $\hat\lambda_{ij}$; using
the same decomposition introduced in the definition of $\rm\bf V$
in (2.9), one explicitly finds:
$$
{\rm\bf A}={1\over2}\left(\matrix{ {\hat\sigma}^* - \hat\eta +2i\hat\omega
& -2\big({\hat\lambda^{(A)}}\big)^*\cr
&\cr
                            -2{\hat\lambda^{(A)}} & \hat\sigma-{\hat\eta}^*- 2i\hat\omega^*}\right)
\qquad
{\rm\bf B}={1\over2}\left(\matrix{ \hat\sigma^* + \hat\eta & -2\big({\hat\lambda^{(S)}}\big)^*\cr
&\cr
                            -2{\hat\lambda^{(S)}} & \hat\sigma+\hat\eta^*}\right)\ ,
\eqno(3.8)
$$
where ${\hat\lambda_{ij}^{(S)}}$ and ${\hat\lambda_{ij}^{(A)}}$ are respectively
the symmetric and antisymmetric parts of the complex matrix $\hat\lambda_{ij}$.

The evolution equation (3.7) has been originally derived using
the equivalent Heisenberg picture where the time-evolution
affects the observables $\cal O$ instead of the states $\rho$;
the two pictures are connected by the duality relation involving
expectation values
$$
\langle {\cal O}\rangle(t)={\rm Tr}\Big[\gamma_t(\rho)\ {\cal O}\Big]
={\rm Tr}\Big[ \rho\ \mit\Gamma_t({\cal O})\Big]\ ;
\eqno(3.9)
$$
the semigroup $\gamma_t$ is generated by (3.1),
while $\mit\Gamma_t$ by the dual equation:
$$
{\partial{\cal O}(t)\over \partial t}=
i\big[H ,{\cal O}(t)\big]+
\sum_{\mu,\nu=1}^{2n} {\rm\bf C}_{\mu\nu} \bigg(
{\rm\bf a}_\mu^\dagger\, {\cal O}(t)\, {\rm\bf a}_\nu
-{1\over2} \Big\{ {\rm\bf a}_\mu^\dagger\, {\rm\bf a}_\nu\, ,\, {\cal O}(t)\Big\}
\bigg)\ .
\eqno(3.10)
$$
Recalling the definition (2.9) and inserting in place of $\cal O$ 
suitable bilinears in the creation and annihilation operators,
one sees that indeed the above equation readily implies (3.7).

Being a first order differential equation involving finite-dimensional
matrices, the evolution equation (3.7) can be easily integrated.
Its solution involves the exponentiation of $\rm\bf A$
and explicitly reads:
$$
{\rm\bf V}(t)= {\rm e}^{t {\rm\bf A}^\dagger}\cdot {\rm\bf V}(0)\cdot {\rm e}^{t {\rm\bf A}}
+\int_0^t{\rm d}\tau\ {\rm e}^{\tau {\rm\bf A}^\dagger}\cdot {\rm\bf B}\cdot {\rm e}^{\tau {\rm\bf A}}\ .
\eqno(3.11)
$$
In the next Sections, the dynamical evolution given by (3.11) 
will be used to investigate whether and under what conditions
it can give rise to entanglement enhancement.

\vskip 1cm

{\bf 4. ENVIRONMENT INDUCED ENTANGLEMENT GENERATION}
\bigskip

In order to study the possibility of entanglement
generation by the external bath, it sufficies
to consider a system formed by two oscillators.
Henceforth, we specialize $n=2$, so that the covariance $\rm\bf V$,
as well as the Kossakowski $\rm\bf C$ and the coefficients
$\rm\bf A$, $\rm\bf B$, all become $4\times 4$ matrices,
and correspondingly $\hat\eta$, $\hat\sigma$ and $\hat\lambda$ 
result $2\times 2$ matrices.
We shall concentrate our attention on discussing bath assisted entanglement
production by purely dissipative mechanism; we shall therefore
ignore any hamiltonian coupling between the two oscillators,
setting in particular $\hat\omega_{ij}=\,0$.%
\footnote{$^\dagger$}{The entanglement power of purely hamiltonian
couplings have been extensively studied in the literature, {\it e.g.} see [6-10].}

Let us first analyze the possibility of entanglement creation at the
beginning of the evolution, in the neighborhood of $t=\,0$.
As initial state of the two-oscillator system 
we shall choose a kernel $\rho(\bar z, z;0)$ representing 
a separable Gaussian state and use the partial transposition criterion [42, 43]
to check whether $\rho(\bar z, z;t)$ becomes entangled
at a later \hbox{time $t$}. 

As well known, in quantum mechanics the operation of full transposition 
on any state $\rho$ corresponds to the
time-reversal transformation: it can be easily implemented in
the holomorphic representation through the variable exchange 
$\bar z_i\leftrightarrow z_i$ in the kernel $\rho(\bar z, z)$.
Recalling the expression in (2.4), this exchange equivalently corresponds
to the transformations $\hat\alpha\leftrightarrow \hat\alpha^*$ and
$\hat\beta\leftrightarrow\hat\beta^T$ on the submatrices defining
the covariance $\rm\bf G$, and thus to the exchange
$a_i\leftrightarrow a_i^\dagger$ in the definition
(2.9) of symmetric covariance $\rm\bf V$. Clearly,
the transformed $\rm\bf V$ still satisfies the positivity
condition (2.12) if the original covariance does, so that
full transposition maps the set of Gaussian density matrices
into itself.

This is not the case for the operation of partial transposition,
{\it i.e.} of transposition involving only one of the two
oscillator system, say the first one, so that $\bar z_1\leftrightarrow z_1$.
It clearly maps states into states for separable ones,
but not in general for correlated ones: it thus provides
a sufficient criterion for bipartite entanglement in any
dimensions. 

For the case at hand, one finds the the operation of partial
transposition with respect to the first system results
in the following transformation of the $4\times 4$ symmetric
covariance:
$$
{\rm\bf V}\rightarrow \widetilde{\rm\bf V}=
{\rm\bf T}\cdot{\rm\bf V}\cdot{\rm\bf T}\ ,
\qquad
{\rm\bf T}=\left(\matrix{ 0 & 0 & 1 & 0 \cr
                          0 & 1 & 0 & 0 \cr
                          1 & 0 & 0 & 0 \cr
                          0 & 0 & 0 & 1}\right)\ .
\eqno(4.1)
$$
We have seen that in order for a Gaussian kernel $\rho(\bar z, z)$ 
in (2.4) to represent a state, the condition (2.12) on its
corresponding symmetric covariance $\rm\bf V$ needs to be
satisfied; if one further finds:
$$
\widetilde{\rm\bf V}+{ {\bf\Sigma}\over 2}<0\ ,
\eqno(4.2)
$$
then the state is surely entangled. In the two-mode case we are studying,
also the converse is true, namely, if $\rm\bf V$ represents the symmetric
covariance of an entangled Gaussian state, 
then (4.2) is necessarily satisfied [44].

In the case of the quasi-free markovian dynamics in (3.11), instead
of dealing directly with the behaviour in time of $\widetilde{\rm\bf V}(t)$
and the inequality (4.2), we find it more convenient
to consider the scalar quantity:
$$
{\cal Q}(t)=\big\langle\Psi\big|\bigg( \widetilde{\rm\bf V}(t)+
{ {\bf\Sigma}\over 2}\bigg) \big|\Psi\big\rangle\ ,
\eqno(4.3)
$$
where $|\Psi\rangle$ is a four-dimensional complex vector,
choosen in the null eigenspace of the matrix \hbox{$\widetilde{\rm\bf V}(0)+{\bf\Sigma}/2$},
so that ${\cal Q}(0)=\,0$. Then, the two oscillators, initially
prepared in a separable Gaussian state, will start to become correlated
by the noisy dynamics induced by the bath in which they are immersed
if a suitable vector $|\Psi\rangle$ exists such that:
$$
\partial_t {\cal Q}(0)<0\ .
\eqno(4.4)
$$
By applying the partial transposition operation $\rm\bf T$ to both
sides of (3.7) and defining:
$$
\big|\widetilde\Psi\big\rangle={\rm\bf T}\, \big|\Psi\big\rangle\ ,
\qquad
\widetilde{\bf\Sigma}={\rm\bf T}\cdot {\bf\Sigma}\cdot {\rm\bf T}
=\left(\matrix{ -\sigma_3 & 0\cr
                  0 & \sigma_3}\right)\ ,
\eqno(4.5)
$$
with $\sigma_3$ the third Pauli matrix, the condition (4.4)
can be cast in the following form:
$$
2\, \big\langle\widetilde\Psi\big| {\rm\bf B} \big|\widetilde\Psi\big\rangle\ <\
\big\langle\widetilde\Psi\big| \Big[ {\rm\bf A}^\dagger \cdot \widetilde{\bf\Sigma}
+ \widetilde{\bf\Sigma}\cdot {\rm\bf A} \Big]\big|\widetilde\Psi\big\rangle\ ,
\eqno(4.6)
$$
explicitly showing the dependence on the bath
through the presence of the coefficients matrices $\rm\bf A$ and $\rm\bf B$.

As initial state for the two oscillators we take a separable state
that is also pure; this is by no means a restriction:
if the bath is not able to entangle
pure states, it will surely not correlate their mixtures.
Gaussian pure states are represented by holomorphic functions
of the form (2.13), where now the matrix $\Omega$ is two-dimensional.
Separability further impose the vanishing of the off-diagonal
elements, so that $\Omega={\rm diag}(\Omega_1, \Omega_2)$,
with $|\Omega_i|\leq 1$ to assure a finite state norm.
Similarly, also the $2\times2$ submatrix $\hat\alpha$ and $\hat\beta$,
defining the covariances $\rm\bf G$ and $\rm\bf V$,
turn out to be  diagonal:
$$
\hat\alpha=\left(\matrix{ {\Omega_1\over 1-|\Omega_1|^2} & 0\cr
                          0 & {\Omega_2\over 1-|\Omega_2|^2} }\right)\ ,
\qquad
\hat\beta=\left(\matrix{ {1\over 1-|\Omega_1|^2} & 0\cr
                          0 & {1\over 1-|\Omega_2|^2} }\right)\ .                    
\eqno(4.7)
$$
The condition (2.12),
$$
{\rm\bf V} +{ {\bf\Sigma}\over 2}=\left(\matrix{ \hat\beta & \hat\alpha\cr
                                                  \hat\alpha^* & \hat\beta^T-1}\right)\geq0\ ,
\eqno(4.8)
$$
that assures the positivity of the corresponding Gaussian state,
gives again the constraints $|\Omega_i|\leq 1$, as it should,
since (4.8) is equivalent to norm convergence.

The null subspace of the combination ${\rm\bf V} + {\bf\Sigma}/2$ in (4.8)
results two-dimensional; it is spanned by the vector
$|\Psi\rangle$ of components $(a\, \Omega_1^*,\ b\, \Omega_2,\ a,\ b)$,
with $a$ and $b$ abitrary complex parameters.
Then, entanglement between the two oscillators will surely occur in baths for which
(4.6) is satisfied with this choice of $|\Psi\rangle$.

To show that indeed this is possible, let us initially prepare the two
oscillators in their corresponding zero-temperature Fock vacua,
so that $\Omega_1=\Omega_2=\,0$. By further choosing the parameters
$a$ and $b$ real and equal, one finds that the inequality (4.6)
reduces to:
$$
\hat\sigma_{11} +\hat\sigma_{22} < {\cal R}e\big(\hat\lambda_{12} +\hat\lambda_{21}\big)\ ,
\eqno(4.9)
$$
involving the entries of the $2\times 2$ matrices $\hat\sigma$ and $\hat\lambda$
that appear in the Kossakowski matrix $\rm\bf C$ in (3.6).
As discussed in Section 3, the matrix $\rm\bf C$ 
parametrizes the physical properties of the bath:
it needs to be positive in order to comply
with the requirement of complete positivity of the reduced dynamics
of the two oscillators. Therefore, in order to be sure that baths satisfying
(4.9) can actually be constructed, one needs to check the compatibility
of (4.9) with the condition of positivity of the Kossakowski matrix.
To explicitly show this, it is enough to take
the matrices $\hat\eta$ and $\hat\sigma$ diagonal and $\hat\lambda$ real,
with only the entry $\hat\lambda_{21}$ non vanishing. Then, the
condition ${\rm\bf C}\geq 0$ simply gives:
$\hat\eta_{11}\geq0$, $\hat\eta_{22}\geq0$, $\hat\sigma_{11}\geq0$,
$\hat\sigma_{22}\geq0$ and $\hat\lambda_{21}^2\leq\hat\eta_{22}\, \hat\sigma_{11}$.
Combining these conditions with that in (4.9), one obtains 
(for $\hat\sigma_{11}\neq0$):
$$
1+{\hat\sigma_{22}\over\hat\sigma_{11}} < {\hat\lambda_{21}\over\hat\sigma_{11}}\leq
\bigg({\hat\eta_{22}\over\hat\sigma_{11}}\bigg)^{1/2}\ .
\eqno(4.10)
$$
As a consequence, in order to be able to correlate the two oscillators, originally
prepared in the separated Fock vacuum,
it is sufficient to immerse them in a bath for which $\lambda_{21}$ is as
in (4.10) and $\hat\eta_{22}$ greater than 
$(\hat\sigma_{11}+\hat\sigma_{22})^2/\hat\sigma_{11}$.

Environments that are able to entangle two oscillators that
are prepared in separable temperature states can similarly be found.
They are characterized by a Kossakowski matrix of the form:
$$
{\rm\bf C}=\left(\matrix{ \eta & \lambda^*\cr
                          \lambda & \sigma}\right)\ \otimes\
           \left(\matrix{ 1 & 1\cr
                          1 & 1}\right)\ ,
\eqno(4.11)
$$
where the parameters $\eta$ and $\sigma$ are real and positive,
while $\lambda$ is complex, satisfying the 
positivity condition: $|\lambda|^2\leq\eta\,\sigma$.
The initial state covariance is characterized by submatrices
$\hat\alpha$ and $\hat\beta$
as in (4.7); for simplicity, let us assume the single
oscillator states to be equal, so that $\Omega_1=\Omega_2=\Omega$.
By taking the norm of the parameters $a$ and $b$ appearing in the vector $|\Psi\rangle$
to satisfy $|a|=|b|=1/\sqrt2$, and further adjusting their phases and that of $\lambda$
according to the relation ${\rm Arg}(\lambda)={\rm Arg}(a)-{\rm Arg}(b)=
\pi-{\rm Arg}(\Omega)$, one checks that the condition (4.6) assuring entanglement generation
can be fulfilled by choosing
$$
|\lambda|> { \sigma-|\Omega|\, \eta\over 1-|\Omega|}\ ,
\eqno(4.12)
$$
which is compatible with the positivity condition of the above Kossakowski matrix
provided we further take $\eta > \sigma$.

\vskip 1cm

{\bf 5. ASYMPTOTIC ENTANGLEMENT}
\bigskip

In the previous Section we have discussed examples of
baths capable of correlate two independent oscillators 
through a purely noisy mechanism: no direct hamiltonian couplings
between the two subsystems were present. This happens at the
beginning of the evolution, as soon as $t>0$. The test
(4.4), on which this conclusion is based, is however
unable to determine the fate of this entanglement
as time becomes large.

On general grounds, one expects that the effects of decoherence,
counteracting entanglement production, be dominant
in the large time region, so that no quantum
correlation is expected to be left at infinity.
Nevertheless, there are situations in which entanglement is
found to keep growing, reaching at the end 
an asymptotic non-vanishing value. An example
is provided by baths for which the corresponding
Kossakowski matrix takes the form in (4.11).

In this case, the evolution equation (3.1) can be
rewritten in a simplified form by introducing
the following set of independent oscillator variables:
$$
\eqalign{ &A={a_1 +a_2 \over\sqrt2} \cr
          &B={a_1 -a_2 \over\sqrt2}  }
\qquad\qquad
\eqalign{ &A^\dagger={a_1^\dagger +a_2^\dagger \over\sqrt2} \cr
          &B^\dagger={a_1^\dagger -a_2^\dagger \over\sqrt2}  \ ,}
\eqno(5.1)
$$
so that: $[A,\, A^\dagger]=[B,\, B^\dagger]=1$ and
all other commutators vanish.
Then, taking into account (4.11), one finds that
the equation (3.1) only involves the operators $A$ and $A^\dagger$:
$$
{\partial\rho\over \partial t}=
-i\omega\big[A^\dagger A ,\rho\big] + L[\rho(t)]\ ,
\eqno(5.2)
$$
where, 
$$
\eqalign{
L[\rho]=\eta\Big(\big[A\rho, A^\dagger\big]+ \big[A,\rho A^\dagger\big]\Big)
&+\sigma\Big(\big[A^\dagger\rho, A\big]+ \big[A^\dagger,\rho A\big]\Big)\cr
&-\lambda \big[A,[A,\rho]\big]
-\lambda^*\big[A^\dagger, [A^\dagger,\rho]\big]\ .}
\eqno(5.3)
$$
For sake of generality, we have also included an hamiltonian
contribution; it comes from an effective Hamiltonian 
of the form (3.3), where, in analogy to (4.11), 
the $2\times 2$ matrix $\hat\omega$ is taken to be:
$$
\hat\omega_{ij}=\omega\ \left(\matrix{ 1 & 1\cr
                          1 & 1}\right)\ ,\qquad \omega >0\ .
\eqno(5.4)
$$
The general solution of the now single-oscillator dissipative evolution
(5.2), (5.3), can be espressed in Gaussian form, using 
the  holomorphic representation involving just one complex variable:
$A^\dagger \rightarrow \bar z$, $A\rightarrow \partial/\partial \bar z$.
It can be explicitly expressed in terms of the analogs of the 
bilinears introduced in (2.6), now just numbers,
$\alpha\equiv\langle A^2\rangle$ and $\beta\equiv\langle A\, A^\dagger\rangle$,
that parametrize the covariance of the corresponding
Gaussian kernel solution $\rho(\bar z,z; t)$.
Notice that we have $\langle A\rangle=\langle A^\dagger\rangle=\,0$
as a consequence of previous choice 
$\langle a_i\rangle=\langle a_i^\dagger\rangle=\,0$.
One finds that the general solution of the evolution 
equations for $\alpha(t)$ and $\beta(t)$
derived from (5.2) is given by [45]:
$$
\eqalign{
&\alpha(t)= e^{-2(\eta-\sigma + i\omega) t}\, (\alpha_0-\alpha_\infty)+ \alpha_\infty\ ,\cr
&\beta(t)=e^{-2(\eta-\sigma)t}\, (\beta_0-\beta_\infty) + \beta_\infty\ ,}
\eqno(5.5)
$$
where the parameters $\alpha_0\equiv\alpha(0)$ and $\beta_0\equiv\beta(0)$ identify 
the initial values, while
$$
\alpha_\infty={\lambda^*(\sigma-\eta+i\omega)\over 
(\eta-\sigma)^2 + \omega^2}\ ,\qquad
\beta_\infty={\eta\over\eta-\sigma}\ .
\eqno(5.6)
$$
An asymptotic equilibrium state exists only when $\eta>\sigma$,
condition that it will henceforth assumed.
Then, the corresponding asymptotic Gaussian kernel is characterized
by a covariance with parameters $\alpha$ and $\beta$ as in (5.6);
it is not a thermal state, unless $\lambda=\,0$,
in which case the temperature $T$ is given by
$\exp(\omega/T)=\eta/\sigma$.%
\footnote{$^\dagger$}{Note that this result does not contradict
the previous conclusions concerning bath-induced entanglement creation,
that in fact requires $\lambda\neq0$.}

On the other hand, all averages involving $B$ and $B^\dagger$ 
result time independent, since (5.2) does not contain any
dependence from these operators; for simplicity, we shall
take them to be all zero, except for the bilinear
$\langle B\, B^\dagger\rangle$:
for reasons that it will appear clear soon,
it is assumed to be equal to $\beta_\infty$,
the long-time asymptotic value of $\langle A\, A^\dagger\rangle$.

Applying back the linear transformations (5.1), these results
can now be used to study the evolution of the true physical 
oscillators, those described by canonical operators $a_i$ and $a_i^\dagger$.
As a result, the two-oscillator state is represented
by a time dependent Gaussian kernel $\rho(\bar z, z;t)$
giving rise to a covariance matrix ${\rm\bf V}(t)$ as in (2.10):
$$
{\rm\bf V}(t)=\left(\matrix{\hat\beta(t) & \hat\alpha(t)\cr
                          \hat\alpha^*(t) & \hat\beta^T(t)}\right) -{{\bf 1}\over 2}\ .
\eqno(5.7)
$$
Its time dependence is encoded into the two functions
defined in (5.5), that appear in the entries of its two
$2\times 2$ submatrices $\hat\alpha_{ij}$ and $\hat\beta_{ij}$;
explicitly, one finds:
$$
\hat\alpha_{ij}(t)={\alpha(t)\over2}\left(\matrix{1 & 1\cr
&\cr
											1 & 1}\right)\qquad
\hat\beta_{ij}(t)={1\over2}\left(\matrix{ \beta(t)+\beta_\infty & \beta(t)-\beta_\infty\cr
&\cr
                            \beta(t)-\beta_\infty & \beta(t)+\beta_\infty}\right)\ .
\eqno(5.8)
$$
For sake of simplicity, we shall further take $\alpha(0)\equiv\alpha_0=\,0$,
so that ${\rm\bf V}(0)$ only involves $\beta_0$ and $\beta_\infty$.
The parameter $\beta_0$ is not completely arbitrary:
the positivity condition (2.12), ${\rm\bf V}(0)+{\bf\Sigma}/2\geq 0$,
readily implies $\beta_0\geq 1$. Note that this two-mode initial state is 
mixed (compare it with the pure case in (4.7)) and further separable;
indeed, after applying the partial transposition operation (4.1), one
finds that
$$
\widetilde{\rm\bf V}(0) +{{\bf\Sigma}\over 2}\geq 0\ ;
\eqno(5.9)
$$
indeed, this condition is equivalent to the inequality
$2\beta_0\beta_\infty\geq \beta_0+\beta_\infty$,
always satisfied for $\beta_0,\ \beta_\infty\geq1$.

We have previously shown that an environment
described by a Kossakowski matrix in the form (4.11)
is able to initially entangle two independent oscillators
immersed in it.
To see whether this still holds in the asymptotic
long-time regime, one needs to examine the properties
of the covariance ${\rm\bf V}_\infty$ 
of the corresponding equilibrium state $\rho_\infty(\bar z,z)$, obtained by
letting $\alpha(t)\rightarrow\alpha_\infty$,
$\beta(t)\rightarrow\beta_\infty$ in the expressions
(5.7), (5.8). By using the partial trace criterion,
$\rho_\infty$ will be entangled if and only if
the $4\times 4$ matrix $\widetilde{\rm\bf V}_\infty +{\bf\Sigma}/2$
posseses negative eigenvalues. In the present case,
these eigenvalues can be analytically evaluated
to be $\beta_\infty \pm \sqrt{\Delta_\pm}/2 -1/2$,
$\Delta_\pm=1+2|\alpha_\infty|^2\pm |\alpha_\infty|(1+|\alpha_\infty|^2)^{1/2}$,
for all four possible combinations of the $\pm$ signs.
Recalling the definitions (5.6), one easily sees that
the lowest eigenvalue 
$\beta_\infty - \sqrt{\Delta_+}/2 -1/2$
can always be made negative
by a suitable choice of the bath parameter $\lambda$.%
\footnote{$^\dagger$}{The actual condition reads:
$|\lambda|^2> 4\eta^2\sigma^2\big[(\eta-\sigma)^2+\omega^2\big]/(\eta^2-\sigma^2)^2$,
which is compatible with the requirement of complete positivity, 
$|\lambda|^2\leq\eta\sigma$, for sufficiently small $\omega$.}

This result is remarkable: it shows that a dissipative
quasi-free dynamics, generated by an equation of the form
(3.1), (3.4), can produce quantum correlations even for large
times, allowing at the end an entangled equilibrium state. 
In view of the
increasing interest that the theory of quantum information
with continous variables is presently attracting, 
these results may be relevant both in phenomenological 
and experimental applications. In particular,
the posssibility of mantaining bipartite entanglement
in a noisy environment even for asymptotic long times
may help the actual realizations of simple quantum
devices in quantum optics and condensed matter physics.

\vskip 1 cm


{\bf REFERENCES}
\bigskip

\item{1.} R. Alicki and K. Lendi, {\it Quantum Dynamical Semigroups and 
Applications}, Lect. Notes Phys. {\bf 286}, (Springer-Verlag, Berlin, 1987)
\smallskip
\item{2.} V. Gorini, A. Frigerio, M. Verri, A. Kossakowski and
E.C.G. Surdarshan, Rep. Math. Phys. {\bf 13} (1978) 149 
\smallskip
\item{3.} H. Spohn, Rev. Mod. Phys. {\bf 53} (1980) 569
\smallskip
\item{4.} H.-P. Breuer and F. Petruccione, {\it The Theory of Open
Quantum Systems} (Oxford University Press, Oxford, 2002)
\smallskip
\item{5.} F. Benatti and R. Floreanini, Int. J. Mod. Phys. 
{\bf B19} (2005) 3063
\smallskip
\item{6.} P. Zanardi, C. Zalka and L. Faoro, Phys. Rev. A {\bf 62} (2000) 030301;
P. Zanardi, {\it ibid.} {\bf 63} (2001) 040304
\smallskip
\item{7.} J.I. Cirac, W. D\"ur, B. Kraus and M. Lewenstein,
Phys. Rev. Lett. {\bf 86} (2001) 544
\smallskip
\item{8.} W. D\"ur, G. Vidal, J.I. Cirac, N. Linden and S. Popescu,
Phys. Rev. Lett. {\bf 87} (2001) 137901
\smallskip
\item{9.} B. Kraus and J.I. Cirac, Phys. Rev. A {\bf 63} (2001) 062309
\smallskip
\item{10.} K. $\dot{\rm{Z}}$yckowski, P. Horodecki, M. Horodecki and R. Horodecki,
Phys. Rev. A {\bf 65} (2001) 012101
\smallskip
\item{11.} D. Braun, Phys. Rev. Lett. {\bf 89} (2002) 277901
\smallskip
\item{12.} M.S. Kim et al., Phys. Rev. A {\bf 65} (2002) 040101(R)
\smallskip
\item{13.} S. Schneider and G.J. Milburn, Phys. Rev. A {\bf 65} (2002) 042107
\smallskip
\item{14.} A.M. Basharov, J. Exp. Theor. Phys. {\bf 94} (2002) 1070
\smallskip
\item{15.} L. Jakobczyk, J. Phys. A {\bf 35} (2002) 6383
\smallskip
\item{16.} Z. Ficek and R. Tanas, Phys. Rep. {\bf 372} (2002) 369
\smallskip
\item{17.} B. Reznik, Found. Phys. {\bf 33} (2003) 167
\smallskip
\item{18.} F. Benatti, R. Floreanini and M. Piani, Phys. Rev. Lett.
{\bf 91} (2003) 070402
\smallskip
\item{19.} F. Benatti and R. Floreanini, Phys. Rev. A {\bf 70} (2004) 012112
\smallskip
\item{20.} F. Benatti and R. Floreanini, J. Opt. B {\bf 7} (2005) S429
\smallskip
\item{21.} F. Benatti and R. Floreanini, Asymptotic entanglement of two
independent systems in a common bath, Int. J. Quant. Inf., 2005, to appear
\smallskip
\item{22.} J.-H. An, S.-J. Wang and H.-G. Luo, J. Phys. A {\bf 38} (2005) 3579
\smallskip
\item{23.} A.S. Holevo, {\it Probabilistic and Statistical Aspects of Quantum Theory},
(North Holland, Amsterdam, 1982)
\smallskip
\item{24.} G. Lindblad, J. Phys. A {\bf 33} (2000) 5059
\smallskip
\item{25.} J. Eisert and M.B. Plenio, Int. J. Quant. Inf. {\bf 1} (2003) 479
\smallskip
\item{26.} B.-G. Englert and K. W\'odkiewicz, Int. J. Quant. Inf. {\bf 1} (2003) 153
\smallskip
\item{27.} A. Serafini, M. Paris, F. Illuminati and S. De Siena,
J. Opt. B {\bf 7} (2005) R19
\smallskip
\item{28.} A. Ferraro, S. Olivares, M. Paris, {\it Gaussian States in Continuous
Variable Quantum Information}, (Bibliopolis, Napoli, 2005)
\smallskip
\item{29.} S.L. Braunstein and P. van Loock, Rev. Mod. Phys. {\bf 77} (2005) 513
\smallskip
\item{30.} B. Demoen, P. Vanheuverzwijn and A. Verbeure,
Rep. Math. Phys. {\bf 15} (1979) 27
\smallskip
\item{31.} P. Vanheuverzwijn, Ann. Inst. H. Poncar\'e {\bf A29} (1978) 123
\smallskip
\item{32.} F.A. Berezin, {\it The Method of Second Quantization},
(Academic Press, Orlando, 1966)
\smallskip
\item{33.} L.D. Faddeev, Introduction to functional methods, in {\it Methods
in Field Theory}, R. Balian, J. Zinn-Justin, eds., (North-Holland, Amsterdam,
1976)
\smallskip
\item{34.} J.-P. Blaizot and G. Ripka, {\it Quantum Theory of Finite Systems},
(The MIT Press, Cambridge, 1986)
\smallskip
\item{35.} A. Perelomov, {\it Generalized Coherent States 
and Their Applications}, (Springer-Verlag, Berlin, 1986)
\smallskip
\item{36.} R. Simon, E.C.G. Sudarshan and N. Makunda, 
Phys. Rev. A {\bf 36} (1987) 3868
\smallskip
\item{37.} R. Simon, N. Makunda and B. Dutta, Phys. Rev. A {\bf 49} (1994) 1567
\smallskip
\item{38.} Arvind, B. Dutta, N. Makunda and  R. Simon,
Pramana, {\bf 45} (1995) 471
\smallskip
\item{39.} V. Gorini, A. Kossakowski and E.C.G. Sudarshan, J. Math. Phys. 
{\bf 17} (1976), 821;
\smallskip
\item{40.} G. Lindblad, Comm. Math. Phys. {\bf 48} (1976) 119
\smallskip
\item{41.} G. Lindblad, Rep. Math. Phys. {\bf 10} (1976) 393
\smallskip
\item{42.} A. Peres, Phys. Rev. Lett. {\bf 77} (1996) 1413
\smallskip
\item{43.} M. Horodecki, P. Horodecki and R. Horodecki, Phys. Lett. {\bf A 223}
(1996) 1
\smallskip
\item{44.} R. Simon, Phys. Rev. Lett. {\bf 84} (2000) 2726
\smallskip
\item{45.} F. Benatti and R. Floreanini, J. Phys. A {\bf 33} (2000) 8139

\bye